\documentclass{IOS-Book-Article}

\usepackage{mathptmx}
\usepackage[final]{listings}
\usepackage{lstsemantic}
\usepackage[all]{xy}
\usepackage[utf8]{inputenc}

\usepackage[T1]{fontenc}
\usepackage[scaled=.9]{beramono}
\usepackage{multicol}
\usepackage{graphicx}
\usepackage{subfigure}
\usepackage{macros}
\usepackage{paralist}
\usepackage{amssymb}
\usepackage{tikz}
\usepackage{url}
\usepackage[show]{ed}
\usepackage{xspace}

\newcommand*{\Hets}{\ensuremath{\mathsf{Hets}}\xspace}
\newcommand*{\DOL}{\ensuremath{\mathsf{DOL}}\xspace}
\newcommand*{\GDOL}{\ensuremath{\mathsf{GDOL}}\xspace}

\newcommand{\technical}[1]{\texttt{#1}}

\newcommand{\ontology}[1]{\technical{#1}}

\def\hb{\hbox to 10.7 cm{}}

\begin{document}

\pagestyle{headings}
\def\thepage{}

\begin{frontmatter}              % The preamble begins here.

%\pretitle{Pretitle}
\title{Extensions of Generic DOL for \\ Generic Ontology Design Patterns}

\markboth{}{Draft June 2019\hb}
%\subtitle{Subtitle}

\author[A]{\fnms{Mihai} \snm{Codescu}%
\thanks{%Corresponding Author: University of Bremen, Collaborative Research Center EASE, Bremen, Germany; E-mail: codescu@uni-bremen.de
This work has been partially supported by the German Research Foundation, DFG, as part of the Collaborative Research Center (Sonderforschungsbereich) 1320 “EASE - Everyday Activity Science and Engineering” %, University of Bremen 
(\url{http://www.ease-crc.org/}).
Corresponding Author's E-mail: codescu@uni-bremen.de
}}
\and
\author[A,B]{\fnms{Bernd} \snm{Krieg-Br\"uckner}}
\and
\author[C]{\fnms{Till} \snm{Mossakowski}}

%\runningauthor{M. Codescu, B. Krieg-Br\"uckner, T. Mossakowski}
%\runningtitle{Generic Ontology Design Patterns in Generic DOL}
\address[A]{University of Bremen, Collaborative Research Center EASE, Bremen, Germany}
\address[B]{German Research Center for Artificial Intelligence (DFKI), CPS, Bremen, Germany % BAALL%, and\\
%University of Bremen, FB3 Mathematik und Informatik, Bremen, Germany
}
\address[C]{Institute for Intelligent Cooperating Systems, Faculty of Computer Science, Otto-von-Guericke University of Magdeburg, Germany}

\begin{abstract}
Generic ontologies were introduced as an extension (%called 
\emph{Generic \DOL}) 
of the \emph{Distributed Ontology, Modeling and Specification Language}, \DOL,  with the aim to 
provide a language for Generic Ontology Design Patterns. In this paper we
present a number of new language constructs that increase the expressivity
and the generality of Generic \DOL, among them
sequential and optional parameters, list parameters with recursion, and
local sub-patterns. These are illustrated with non-trivial patterns: 
generic value sets and (nested) qualitatively graded relations, 
demonstrated as definitional building blocks in an application domain.
\end{abstract}

\begin{keyword}
ontology design patterns\sep
recursive pattern definition\sep
generic ontologies\sep
generic DOL\sep
qualitatively graded relations
\end{keyword}
\end{frontmatter}
\markboth{}{}

\section{Introduction}\label{sec:intro}

Ontology design patterns (ODPs) \cite{DBLP:books/ios/HGJKP2016} 
have been introduced as a means to establish best practices for ontology
design as well as a way to provide a set of carefully-designed
building blocks for ontologies that can be reused in different contexts.
In \cite{DBLP:conf/semweb/Krieg-BrucknerM17}, we have introduced
\emph{generic} ontology design patterns (GODPs), using the language 
Generic \DOL. While simple ODPs usually are just ontologies, GDOPs
have \emph{parameters} that can be \emph{instantiated} in different
ways, thus leading to an even greater and easier re-use of ODPs.

This paper is an update of \cite{DBLP:conf/semweb/Krieg-BrucknerM17}, addressing many of the extensions of Generic \DOL listed
there as future work, on which we focus in this paper. 
To better illustrate the degree of generality
provided by the new language extensions,
we deliberately decided %chose here 
to present some of the examples used in the cited paper
again in their new form.

Further examples of GODPs formulated in Generic \DOL{} can be found in \cite{GDOL-BOG19},
in particular a role pattern from the literature, reformulated in a modular and
reusable way. 
\cite{GDOL-BOG19} also gives more motivation and describes the advantages of Generic ODPs over ``classical'' ODPs,
i.e. of parameterization over subsumption (parametric polymorphism over subtype polymorphism, resp.).

%\ednote{TODO. @BKB, @TM: could you add something here?}

%\ednote{Should we cite our Jowo-BOG submission ``Generic Ontology
%  Design Patterns at Work''? MC: I think it would help if each of the papers
%  cited the other and explains a bit the relationship. TM: done (at
%the end of the paper).}

%\ednote{In the BOG paper, we have replaced the keyword \textbf{ontology}
%by \textbf{pattern} in the case of (G)ODPs. Should we do the same here?
%MC: we can do this, actually I proposed this too. 
%BKB also suggested \textbf{generic}. TM: replaced it with \textbf{pattern}.}

\section{Generic \DOL}\label{sec:gdol}

\lstset{% general command to set parameter(s)
basicstyle=\small, % print whole listing small
basicstyle=\ttfamily,
%keywordstyle=\color{black}\bfseries\underbar, % underlined bold black keywords
%identifierstyle=, % nothing happens
%commentstyle=, % white comments
stringstyle=\sffamily, 
showstringspaces=false,
%language={},
numbers=left, numberstyle=\tiny, stepnumber=2, numbersep=5pt,
language=dolText,alsolanguage=owl2Manchester}

The \emph{Distributed Ontology, Modeling and Specification Language}, %shortly
 \DOL \cite{mossakowski2015distributed}, 
is a meta-language that enables modular development of ontologies\footnote{We
ignore the modeling and specification aspects here and restrict to 
ontologies only.} and allows specification of intended relationships
(e.g. theory interpretation, alignment, properties of extensions) between them. 
\DOL is supported by the \emph{Heterogeneous Tool Set}, %shortly
\Hets \cite{conf/tacas/MossakowskiML07}, that provides a parser for \DOL specifications,
an implementation of \DOL semantics, and an interface to theorem provers.

The language \emph{Generic \DOL} (or \GDOL) was proposed in
\cite{DBLP:conf/semweb/Krieg-BrucknerM17} as an extension of \DOL 
with parameterized ontologies (that we may also call generic ontologies), 
following generic specifications in CASL \cite{CASL-RM}. 
Parameters of generic ontologies are
ontologies themselves: 
thus we may require that some abstract properties 
are expected to hold
for the argument ontology provided in an instantiation.
This provides important semantic expressivity, going 
beyond macro approaches such as OTTR \cite{DBLP:conf/semweb/SkjaevelandKL18}, 
where parameters are just symbols of a certain kind, or lists of
these. Moreover, a generic ontology may import ontologies, 
written after the list of parameters with the keyword \texttt{given}.
The semantics is that the symbols of these ontologies are visible in the
parameters and the body, but will not be instantiated.

\begin{figure}[t]
  \begin{multicols}{2}
\begin{lstlisting}[firstline=2]
pattern ReflexiveRelation 
 [ObjectProperty: r]
 [Class: C] =
 ObjectProperty: r  
  Domain: C  Range: C
  Characteristics: 
    Reflexive
\end{lstlisting} 
\begin{lstlisting}[firstline=2]
pattern InverseRelation
 [ObjectProperty: r]
 [Class: D][Class: R]
 [ObjectProperty: ir] =
 ObjectProperty: r  
  Domain: D  Range: R
 ObjectProperty: ir  
  Domain: R  Range: D
  InverseOf: r
\end{lstlisting}  
    \columnbreak
\begin{lstlisting}[firstline=2]
pattern TransitiveRelation 
 [ObjectProperty: r]
 [Class: C] =
 ObjectProperty: r  
  Domain: C  Range: C
  Characteristics: 
    Transitive  
\end{lstlisting}
\begin{lstlisting}[firstline=2]
pattern SubProp
 [ObjectProperty: q]
 [Class: D][Class: R]
 [ObjectProperty: p
  Domain: D  Range: R] =
 ObjectProperty: q  
  Domain: D  Range: R
  SubPropertyOf: p    
\end{lstlisting} 
    \end{multicols}
    \caption{\ontology{ReflexiveRelation}, \ontology{TransitiveRelation},
    \ontology{InverseRelation} and \ontology{SubProp}.}
    \label{fig:simple-patterns}
\end{figure}

Several simple examples of generic ontologies are
presented in Fig.~\ref{fig:simple-patterns}, where we introduce patterns for
very basic building blocks for ontologies:
\ontology{ReflexiveRelation} adds, for a given object property \texttt{r}
and a given class \texttt{C},
axioms stating that  \texttt{r} is reflexive, and its domain and range are \texttt{C};  
\ontology{TransitiveRelation} adds domain, range and transitivity
axioms, given one object property and one class; 
\ontology{InverseRelation} takes two object properties and two classes, and
adds domain and range axioms for them, and an axiom that one object
property is the inverse of the other.
The ontology \ontology{SubProp} already makes use of a new language
construct of Generic \DOL %that we introduce in this paper; it will be explained 
(cf. the next section):
%For the first three generic ontologies,
%the parameters contain only declarations of symbols, but \ontology{SubProp} 
it has, as fourth parameter, an ontology that contains two additional axioms.

Just like for OWL (and \DOL), the 
\emph{``Same Name -- Same Thing'' principle} is used, which means that
the definition of an entity can be repeated without introducing multiple
occurrences of that entity. For Generic \DOL, this means that if the body
of a generic ontology declares an entity, the union of multiple instantiations
of that generic ontology will contain only one occurrence of that entity. 
If this was not the intention, the entity should rather become a 
parameter of the generic ontology, such that each instantiation can
assign it a different name.

\section{Extensions of Generic \DOL} \label{sec:gdol-ext}

\subsection{Sequential Parameters, Local Environments, and Compact Notation for Arguments}

%\ednote{BKB: perhaps this paragraph should go down to the explanation with examples.
%MC: done. I first put it in the previous section because it is not a 
%contribution of this paper.}
%A methodological decision is 
It is in the interest of simplicity (of writing and reading)
to keep the parameters of generic ontologies 
as small as needed. The aim is to avoid having to explicitly provide 
symbol mappings when making instantiations of generic ontologies: if a parameter 
and its corresponding argument 
consist each of only one symbol, \Hets will automatically derive 
the unique way of mapping the one to the other. When such derivations cannot be done
automatically, the user must specify how symbols are mapped using a 
sequence of mapping items, written \texttt{f |-> a} where \texttt{f} and \texttt{a}
are symbols of the parameter and argument, respectively. 
The sequence of mapping items gives rise to a morphism between the 
parameter and the argument, called \emph{fitting morphism}.
To make the notation
more compact, the parameters of a generic ontology 
and the arguments in an instantiation
may be written in Generic \DOL{} as a
semi-colon separated list, e.g. 
%\texttt{TransitiveRelation[ObjectProperty: r; Class: C]}
%instead of 
%\texttt{TransitiveRelation[ObjectProperty: r][Class: C]}
%and
\texttt{TransitiveRelation[olderThan;Person]}
instead of
\texttt{TransitiveRe\-lation[olderThan][Person]}.%
\footnote{For OWL we can use semicolons as separators between parameters and
arguments because they are not used as separators between declarations
at the basic level. When this is the case, e.g. for the CASL logic,
where one can write \texttt{sort s; op c : s}, we will use  curly
braces to mark the beginning and the ending of an ontology. Thus,
\texttt{G[sort s; sort t]} will be parsed as
\texttt{G[\{sort s; sort t\}]} -- an ontology declaring two sorts,
while a generic ontology with two sorts as parameters or arguments
will be written
\texttt{H[\{sort s\};\{sort t\}]}.  
}

The first significant extension of Generic \DOL{} that we introduce is a modification of the
semantics of generic ontologies. In the semantics of generics in
\cite{DBLP:conf/semweb/Krieg-BrucknerM17,CASL-RM}, each parameter forms its 
own environment, and sharing between parameters is allowed only
if the image of each shared symbol through the fitting morphisms of the
parameters where it occurs is the same. \Hets checks this and issues an error
message when this condition does not hold for an instantiation. In the
context of %our decision of 
keeping the parameters small,
we decided to allow each parameter to share the environment of all
parameters preceding it along a chain of inclusions. 
We call this \emph{sequential semantics} for parameters of generic ontologies.

As an example, the generic ontology \ontology{SubProp} in Fig.~\ref{fig:simple-patterns} takes as parameters an object property \texttt{q},
two classes \texttt{D} and \texttt{R}, and finally an ontology 
extending the previous parameters with the declaration of an object property
\texttt{p} with domain \texttt{D} and range \texttt{R};
 its body adds axioms that domain and range of \texttt{q} are
also \texttt{D} and \texttt{R}, respectively, and moreover
\texttt{q} is a sub-property of \texttt{p}.
Note that with the sequential semantics it has become possible to refer to  \texttt{D} and  \texttt{R} in the axioms for \texttt{p};
the effect of including the domain and range axioms for the parameter \texttt{p}
now allows (indeed requires) the checking of these as constraints on its argument in each instantiation, as we shall see below.

The semantics of instantiation in
\cite{DBLP:conf/semweb/Krieg-BrucknerM17,CASL-RM}
imposes a compatibility condition between the fitting morphisms
for the different parameters: if a symbol occurs in multiple formal
parameters, it must be mapped by the different fitting morphisms
in a unique way. This compatibility condition
remains in our extension, and the user can rely on it to provide  
symbol mappings only for the 
new symbols of a parameter; by compatibility, the
way the old symbols are mapped is already defined.
To illustrate this, let us assume we want to define the 
\texttt{isAncestorOf} property
between \texttt{Person}s as a transitive relation and with subproperty
\texttt{isParentOf} (cf. Fig.~\ref{fig:ancestor}). 
We first instantiate the pattern \ontology{TransitiveRelation} 
to obtain that \texttt{isAncestorOf} is transitive and has
domain and range \texttt{Person}. We would like
to write \texttt{[isAncestorOf]} as a shorthand notation for the
argument
\texttt{TransitiveRelation[isAncestorOf;Person;Person]}, or,
to be fully correct, as this ontology has more than one object property,
as a shorthand notation for
the even longer form where this is followed by 
\texttt{fit p |-> isAncestorOf}.

\begin{figure}[h]
\begin{lstlisting}[morekeywords={let,in,given,EquivalentWith}]
ontology PersonRels = 
 TransitiveRelation [isAncestorOf; Person]
then SubProp[isParentOf; Person; Person; isAncestorOf]
\end{lstlisting} 
    \caption{Using the patterns for a concrete design.}
    \label{fig:ancestor}
\end{figure}

This requires another language extension regarding instantiation of
parameters of generic ontologies. Firstly, in the case of %\DOL extensions
some \DOL{} ontology  \texttt{O1}
%\ednote{BKB=>MS: reformulation OK here? MC: I changed text to ontology}
%\ednote{BKB=>MS: actually, this is the ``given''; have you explained this here and in the other paper? MC: No, this is not the ``given''. Given only makes sense for generic ontologies. If the ontology is not generic, as it is the case here,
%the local environment is what comes before the ``then''. Given was explained at some point, but I see now that the explanations is missing. I add it back here and will do so in the other paper, but we don't have any example in this paper where this is used.}
 followed by an instantiation, written \texttt{O1 then G[AP1]}, the 
\emph{local environment} \texttt{O1} of previous declarations that is being 
extended is implicitly added to the argument, i.e. this expands to
\texttt{O1 then G[O1 then AP1]}. In the case of ontologies with imports,
the local environment of an instantiation will include them.
Secondly, we introduce a shorthand notation
for the instantiation of those parameters that define only one new symbol 
(recall that we assume sequential semantics of  parameters, thus
the symbols of all previous parameters are visible at each step). Consider that
the name of this unique new symbol of a parameter is \texttt{N} and
its kind (class, object property, etc.) is $k$. For an
instantiation of that parameter, in 
\cite{DBLP:conf/semweb/Krieg-BrucknerM17,CASL-RM} an ontology is
required as an argument, which can be given in two forms:
\begin{itemize}
\item as a named ontology \texttt{O}. Then we must be able to derive uniquely how \texttt{N} is mapped to a symbol of kind $k$ in 
\texttt{O}, otherwise we must explicitly provide  a symbol mapping of the form
\texttt{N |-> N'} where \texttt{N'} is a symbol of kind $k$ in \texttt{O}.
\item %an explicitly kinded symbol
  as an anonymous ontology consisting of a sequence of symbol declarations
  and axioms. A special case is that of a single symbol defined with an
  explicit kind; then this kind must also be $k$. In such a case, this unique symbol 
  is considered newly declared and acts as an argument, 
  and the symbol mapping is uniquely determined.
\end{itemize}
We here propose a third option:
\begin{itemize}
\item the name \texttt{M} of a symbol of kind $k$ from the local
environment is passed as an argument. 
The argument expands to \texttt{E fit N |-> M}. Thus any properties
that \texttt{N} must have, as specified in the parameter, are checked
for the symbol \texttt{M} in the local environment. 
\end{itemize}

In general, an instantiation \texttt{SubProp[sr;A;B;r]} of \ontology{SubProp}
where \texttt{A}, \texttt{B} are classes and \texttt{sr}, \texttt{r} are
relations from the local environment, can be done only, if \texttt{r} has domain
\texttt{A} and range \texttt{B}. The result is that \texttt{sr} becomes a 
sub-property of \texttt{r} and moreover it gets 
domain \texttt{A} and range \texttt{B} (if this was not already available in the
local environment). An instantiation \texttt{SubProp[sr;A;B;r]}
where \texttt{sr} is not visible in the local environment
also requires that \texttt{r} has domain
\texttt{A} and range \texttt{B}. The result is that a new relation \texttt{sr}
 is defined and added to the local environment
(again, with domain \texttt{A} and range \texttt{B}, and as a sub-property
of \texttt{r}).

In our example, we can then write \texttt{isAncestorOf} as an argument for the
fourth parameter of \texttt{SubProp}; with the third case listed above,
this means that we refer to the symbol declared in the instantiation of
\texttt{TransitiveRelation}, and therefore
the expected domain and range axioms hold for \texttt{isAncestorOf}.

We make use of the simple patterns defined
in Fig.~\ref{fig:simple-patterns} to extend a simple order (the transitive 
\texttt{greater[Val]} relation on a class \texttt{Val}):  in Fig.~\ref{fig:order-rel-ext}
%:\ednote{TM: the instantiations in Fig.~\ref{fig:order-rel-ext}
%  are already difficult to read... MC: one solution would be to write another pattern for the case when the domain and the range are the same, then we have fewer
%  arguments. Should I do that? TM: for TransitiveRelation, I would do
%  this anyway, but maybe not for the others. Do we really need parameterised
%  names here? Could we rename Val into V? New idea: separate lists of
%arguments with semicolons.}
we define its inverse \texttt{less[Val]}, 
a \texttt{greaterOrEqual[Val]} and a \texttt{lessOrEqual[Val]} relation 
that are inverse to each other
such that \texttt{less[Val]} is a sub-property of \texttt{lessOrEqual[Val]}
and \texttt{greater[Val]} is a sub-property of \texttt{greaterOrEqual[Val]}.

\paragraph{Parameterized Names} 
Here the symbols declared in the two patterns have  \emph{parameterized names},
to make explicit that they depend on the names of the parameters. The notation for
parameterized names is \texttt{Name[Param1, ..., ParamN]}, if the name of the 
new symbol depends on \texttt{N} parameters. During instantiation, the names of the
arguments are substituted in the parameterized name, e.g. 
\texttt{greater[Val]} becomes \texttt{greater[Significance]} if the value 
provided for \texttt{Val} is \texttt{Significance}. \Hets also offers the 
possibility of \emph{stratifying} these names for the result of an instantiation:
the name \texttt{greater[Significance]} is replaced with \texttt{greater\_Significance}, thus obtaining a legal OWL identifier. 
\begin{figure}[h]
\begin{lstlisting}[firstline=2]
pattern SimpleOrder [Class: C] =
 TransitiveRelation[greater[C]; C] 

\end{lstlisting}
\begin{lstlisting}[firstline=2]
pattern OrderRelationExtension
 [Class: Val;
  TransitiveRelation[greater[Val]; Val] ] =
     InverseRelation[less[Val];          Val;Val;greater[Val]] 
 and InverseRelation[greaterOrEqual[Val];Val;Val;lessOrEqual[Val]]
 and ReflexiveRelation[lessOrEqual[Val]; Val] 
then TransitiveRelation[less[Val];           Val] 
 and TransitiveRelation[greaterOrEqual[Val]; Val] 
 and TransitiveRelation[lessOrEqual[Val];    Val] 
 and SubProp[greater[Val]; Val; Val; greaterOrEqual[Val]] 
 and SubProp[less[Val];    Val; Val; lessOrEqual[Val]]
\end{lstlisting} 
    \caption{\ontology{SimpleOrder} and \ontology{OrderRelationExtension}.}
    \label{fig:order-rel-ext}
\end{figure}

 As an argument of
 \ontology{OrderRelationExtension}
 we could provide any transitive relation, in particular,  a
 strict order. Since OWL does not support transitive and asymmetric relations,
 the argument would have to be given in a logic where this can be expressed, 
 e.g. OWL with restrictions \cite{DBLP:conf/owled/SchneiderRS13} or
 first-order logic. The theory presented informally in this paper is actually
 independent of the underlying formalism used for writing ontologies 
 (OWL in the examples here) and moreover provides 
 support for heterogeneous specifications as in the above example:
 the parameter can be instantiated with an argument in another logic
 along an encoding of the logic of the parameter
 to the logic of the argument.

\subsection{Local Sub-Patterns, Optional Parameters, List Parameters, and Recursion}
%\ednote{BKB=> Mihai: I always use capitals in section headers; can you check what the editors require? MC: I change to capitals.}

A pattern can be structured into smaller sub-patterns; often we want to 
make these visible only in the pattern where they are introduced.
For this, we allow \emph{local definition of sub-patterns} before the body of a
generic ontology, using a \texttt{let} notation. The local sub-patterns
share the parameters of the main pattern where they are defined. 
Note that this considerably abbreviates the notation; 
in effect, it corresponds to a partial instantiation of a corresponding pattern 
declared outside of the body (cf. \cite{DBLP:conf/semweb/Krieg-BrucknerM17}).
The body of the main pattern may, and in most cases will, make use of 
instantiations of the local sub-patterns.

We may mark parameters as \emph{optional}, written \texttt{?[FP]} (as in OTTR \cite{DBLP:conf/semweb/SkjaevelandKL18}),
where \texttt{FP} is a parameter, or 
\texttt{[ ...; ? FP; ...]} in the notation with semicolons. At instantiation, if an argument 
is not provided for an optional parameter (written \texttt{[]} or 
as a whitespace between semicolons \texttt{; ;}), 
all occurrences of that parameter in the body are replaced with the 
empty ontology, and
all symbols and sentences containing symbols from that parameter are removed.
%\ednote
%{TM: I have the impression that the formal semantics of this is
%  quite involved. Note that the semantics of the body is given in
%  a form where the reference to \texttt{FP} has already been resolved.
%  We could remove all symbols and axioms from \texttt{FP} using
%  \texttt{reject}, but this may remove too much, namely in cases
%  where symbols occuring in \texttt{FP} have also been declared
%  elsewhere in other parameters or in the body. 
%  Hence, in order to give a semantics to ``all occurrences of that parameter
%  in the body are replaced with the empty ontology'', it seems that
%  we need to re-compute the semantics of the body in a global
%  environment where \texttt{FP} is bound to the empty ontology.
%  But this generally will not be well-formed, because the body may contain
%  declarations (symbols and axioms) depending on \texttt{FP},
%  which shall be removed. Hence, we would need a ``remove things that
%  are not well-formed'' mode of the semantics. Ugghhh...
%  MC: I was also thinking of re-computing the semantics of the body.
%  I think it would work if we reject from the old body 
%  those symbols from the optional parameters 
%  that don't appear in other parameters, in an env where \texttt{FP} is
%  empty ontology if it's a missing optional parameter. TM: OK,
%  maybe this works.}

We also introduce language constructs for \emph{list parameters}, 
in spirit similar to those in OTTR \cite{DBLP:conf/semweb/SkjaevelandKL18}.
While OTTR patterns support only iteration and zip over list parameters,
we allow recursive calls of patterns over lists in Generic \DOL, which would
be considered illegal in OTTR because they introduce cyclic dependencies 
between patterns.
%\ednote{TM: how do we go beyond OTTR's recursion?
%MC: what I understand from the OTTR paper cited here and the OTTR homepage,
%recursion there refers to the way the pattern is built by replacing
%instances. With list parameters you can do iteration (cross mode) and zip.
%We have recursive patterns in the sense that their body refers to a call
%of the same pattern on a smaller argument list. In OTTR no cyclic dependencies 
%are allowed, and I think our recursive patterns qualify as such.
%TM: Aha, so OTTR only supports some fixed patterns of iteration, while
%we support full recursion. Maybe stress this more? MC: done, please check and improve if needed.} 
A list is written \texttt{X :: Xs}, where \texttt{X} is an ontology
and \texttt{Xs} denotes the tail of the list. 
If \texttt{X} is an ontology declaring only one symbol of a certain
kind, it is assumed that all the ontologies \texttt{Xs} are of the same form. 
We may refer to such list as a list of symbols of that kind.
For example, \texttt{Class: C :: Cs} is a list of ontologies each consisting
only of a class declaration.
An ontology with such a  list as a parameter
is written 
\texttt{ontology G [Class: C :: Cs] = ...}. The
empty list is written \texttt{[empty]} and is treated as an empty optional
argument.

\paragraph{Notations.} In the argument
of an instantiation of a generic ontology \texttt{G}, we may %also
write
\begin{itemize}
\item \texttt{[]} for \texttt{[empty]},
\item \texttt{[X]} for \texttt{[X::empty]}, and
\item \texttt{[X$_1$, $\ldots$, X$_n$]} for
\texttt{[X$_1$ :: $\ldots$ :: X$_n$ :: empty]}.
\end{itemize}

\paragraph{Value Sets.}
Qualitative values, corresponding to abstractions from quantitative data, occur quite often in practice, cf. grading below. 
As we know from cognitive science, they are related to the human need for doing away with irrelevant detail (precision in this case); here (and there) they allow us to simplify abstract reasoning (cf. \cite{DBLP:conf/popl/CousotC77}). 

%\ednote{TM: please first motivate and explain this pattern. @BKB: could you please do this?}
With the new constructions introduced above, 
the pattern \texttt{ValSet} (Fig.~\ref{fig:val-set})
%  TM: I have provided a modified and simplified version, see 
%   Fig.~\ref{fig:fin-lin-order}. This can be explained much more easily. I have used a list of
%  individuals (not classes) as parameters, which should simplify the
%  things considerably and also is ontologically more valid. 
%  MC: we followed your suggestion, modulo some rewriting. I would still call this
%  ValSet and keep the order optional, because we want to illustrate this 
%  GDOL language feature.
 has as 
arguments: a class of values, a list of value individuals, and an optional 
relation between these values.
The sub-pattern \texttt{OrderStep} introduces the fact that a value belongs to the
set of values and is optionally greater than the value introduced in the
set at the previous step. 
%The sub-pattern \texttt{SingleVal} introduces the only individual \texttt{VAL[C]}
%of a class \texttt{C} that is a subclass of \texttt{Val}.
%The sub-pattern \texttt{GrVal} introduces the fact that one 
%individual is greater than the other.
%The sub-pattern \texttt{Step} iterates by recursion over a list of classes: 
%at each step it defines a new value that is greater than the one obtained at 
%the previous step, which is passed as an extra parameter. 
Once the list \texttt{vS} is empty, the recursion stops.
All this is put together in the body of \ontology{ValSet}:
the value is created for the first element of the list of value individuals,
the relation \texttt{greater} is defined to be a simple order on
\texttt{Val},
the iteration creates the rest of the values, and finally
the values are declared to be different from each other and
the set of values is defined to be the disjoint union of all values.

\begin{figure}[t] 
%\begin{multicols}{2}
%    \columnbreak
\begin{lstlisting}[morekeywords={let,in,given,EquivalentWith},firstline=2]
pattern ValSet [Class: Val; Individual: v0 :: vS;
                 ? ObjectProperty: greater       ] 
= %% all individuals vi from v0::vS become members of Val
let OrderStep [Individual: vi; Individual: vj :: vS] 
    =  Individual: vj Types: Val Facts: greater vi
     then OrderStep[vj; vS]
    end
in   Individual: v0 Types: Val
then SimpleOrder[greater;Val] and OrderStep[v0; vS]
then { DifferentIndividuals: {v0 :: vS}
       Class: Val EquivalentWith: {v0 :: vS} } 
\end{lstlisting} 
%\end{multicols}    
    \caption{\ontology{ValSet}.}
    \label{fig:val-set}
\end{figure}

%\begin{figure}[t] 
%\begin{multicols}{2}
%%    \columnbreak
%    \lstinputlisting[morekeywords={let,in,given,EquivalentWith}]{Examples/FiniteLinearOrder.dol} 
%\end{multicols}    
%    \caption{\ontology{Finite linear order}.}
%    \label{fig:fin-lin-order}
%\end{figure}

The optional parameter for \ontology{ValSet} allows to create instances of this 
pattern both for the case when the values are ordered
(\ontology{ValSet\_Significance} in Fig.~\ref{fig:val-set-insts}),
%where \texttt{O} provides the required domain and range axioms for
%\texttt{greater[Significance]},
and for the case when the values in the set are not ordered 
(\ontology{ValSet\_CrustStyle} in Fig.~\ref{fig:val-set-insts}).
The expansion of \ontology{ValSet\_Significance} is precisely
the ontology \ontology{GradedRelations4Exp} in Fig.~3 of \cite{DBLP:conf/semweb/Krieg-BrucknerM17}.
%\ednote{@TM: this should have been called \ontology{SignificanceExp}
%in that paper, agreed?. TM: yes} 
We may also extend the order relation \texttt{greater[Val]} on the
value set with its inverse \texttt{less[Val]}, its reflexive version
\texttt{greaterOrEqual[Val]} and the inverse of its 
reflexive version \texttt{lessOrEqual[Val]},
as illustrated in Fig.~\ref{fig:val-set-with-order}.

\begin{figure}[t] 
\begin{multicols}{2}
\begin{lstlisting}[morekeywords={let,in,given,EquivalentWith}]
ontology ValSet_CrustStyle =
 ValSet[CrustStyle;
        [bottomCrust,
         topCrust,
         singleCrust,
         twoCrust,
         turnoverCrust,
         strudelCrust ];
        %% no order
       ]
\end{lstlisting} 
    \columnbreak
\begin{lstlisting}[morekeywords={let,in,given,EquivalentWith}]
ontology ValSet_Significance =
 ValSet[Significance;
        [0Insignificant,
         1Subordinate,
         2Essential,
         3Dominant     ];
        greater[Significance] ]
\end{lstlisting} 
\end{multicols}    
    \caption{Instantiations of \ontology{ValSet}.}
    \label{fig:val-set-insts}
\end{figure}

\begin{figure}[t] 
\begin{lstlisting}[morekeywords={let,in,given,EquivalentWith}]
pattern ValSetWithOrder[Class: Val; Individual: v :: Vs] =
 OrderRelationExtension[Val; ValSet[Val; v::Vs; SimpleOrder[Val]]]
\end{lstlisting} 
    \caption{Extending the order on \ontology{ValSet}.}
    \label{fig:val-set-with-order}
\end{figure}

\paragraph{Graded Relations.}
In \cite{DBLP:conf/semweb/Krieg-BrucknerM17} we introduced 
a pattern for graded relations with a grade domain with 4 values
and stated that analogous patterns must be provided for each number of values.
The main idea of the pattern \cite{DBLP:conf/ksem/Krieg-Bruckner16} 
is to introduce a qualitative metric,
arbitrarily fine and usually represented as an ordered set, for an object property.
Typical examples include the significance of an ingredient in a recipe,
or how much a person is affected  by an impairment. Instead of using
reification for the ternary relation thus obtained, the solution
proposed in \cite{DBLP:conf/ksem/Krieg-Bruckner16} is to encode the grading
with a sheaf of relations, one for each grade. The intended meaning is that 
\begin{center}
\texttt{hasTarget(?s,?t,Val)} $\equiv$ \texttt{hasTarget\_Val(?s,?t)}
\end{center}
\noindent for a ternary relation \texttt{hasTarget} with grade value \texttt{Val}
as third argument. 

Using list parameters and recursive sub-patterns, 
we can now provide one pattern that covers all %possible
necessary numbers of values, as in Fig.~\ref{fig:graded-rels}.
The last parameter of \ontology{GradedRels} is a list of 
ontologies, with the assumption that each of them declares an individual
of type \texttt{Val}. The local sub-pattern \ontology{Step} has as
parameter a list of ontologies such that each of them declares an individual.
In the instantiation \texttt{Step[Val::ValS]}, the first element of the 
argument list is the ontology obtained by expanding the notation \texttt{Val}, i.e.
the local environment, in this case the union of all formal parameters
that we denote \texttt{Env},
contains a declaration for the individual \texttt{Val}. The argument 
expands then to \texttt{Env fit X |-> Val}. By assumption, each element of
the list of ontologies \texttt{valS} declares an individual (and an axiom 
about its type, that is not needed here),  so we can
use it as an argument for \texttt{xs}, which is a list of ontologies 
each declaring an individual.
% \ednote{@MC: explain that arguments of parameterised ontologies are
%  ontologies themselves, and the magic MC: done, please check.}

%\ednote{TODO: explain 
%  more if enough space. TM: I think without more explaination this
%is hard to digest. MC: done, please check.} 
%\ednote{Fig.~\ref{fig:graded-rels} uses an name \texttt{with}, but this
%is already a keyword in \DOL{}. MC: replaced to \texttt{withGrade} until we 
%get a better idea. MC: agreed to modify compound ids such that
%\texttt{p[X]} would get mapped to \texttt{r[C]} instead of \texttt{p[C]}.}

\begin{figure}[t] 
\begin{lstlisting}[morekeywords={let,in,given,EquivalentWith}]
pattern GradedRels 
  [Class: S; Class: T; 
   ObjectProperty: p Domain: S Range: T; 
   Class: Val;
   {Individual: v Types: Val} :: valS  ]
= %% a sheaf of graded relations p[vi], one for each vi in v::valS
let Step[Individual: x :: xs] =
    { ObjectProperty: p[x] Domain: S Range: T SubPropertyOf: p }
     then Step[xs]
    end
in Step[v::valS] and 
   { ObjectProperty: has[Val] Domain: T Range: Val }
\end{lstlisting} 
\begin{lstlisting}[morekeywords={let,in,given,EquivalentWith}]
ontology GradedRels_Significance =
 GradedRels[PhysicalObject;PhysicalObject;hasIngredient;
  Significance;[0Insignificant,1Subordinate,2Essential,3Dominant]]
\end{lstlisting} 
     \caption{\ontology{GradedRels} and instantiation \ontology{GradedRels\_Significance}.} 
    \label{fig:graded-rels}
\end{figure}

%\begin{figure}[t] 
%    \lstinputlisting[language=dolText,mathescape,morekeywords={Class, ObjectProperty,Domain,Range,SubPropertyOf,given,fit,let,in,Characteristics,
%    Reflexive,InverseOf,Transitive}]{Examples/GradedRelsLE_SignificanceExpanded.dol} 
%     \caption{Expansion of \ontology{GradedRelsLE\_Significance}.} 
%    \label{fig:graded-rels-expanded}
%\end{figure}

\subsection{Template Matching for List Parameters}

%\begin{figure}[t]
% \lstinputlisting[morekeywords={let,in,given,EquivalentWith}]{Examples/GradedRelsGE.dol}
%    \caption{\ontology{GradedRelsGE}.} 
%    \label{fig:graded-rels-le}
%\end{figure}

\begin{figure}[t]
%\begin{multicols}{2}
\begin{lstlisting}[morekeywords={let,in,given,EquivalentWith}]
pattern GradedRelsSub  
 [Class: S; Class: T; ObjectProperty: p Domain: S Range: T; 
  Class: Val; {Individual: v0 Types: Val}
               :: {Individual: v1 Types: Val} :: valS]  =
let
 Sub2[ObjectProperty: r; ObjectProperty: r1; ObjectProperty: r2] =
  ObjectProperty: r  Domain: S Range: T
  ObjectProperty: r1 SubPropertyOf: r
  ObjectProperty: r2 SubPropertyOf: r
 end 
 AtMostInitial[Individual: x; Individual: y :: empty] =
  Sub2[p[atMost[y]];  p[x];          p[y]]
 end
 AtMostStep [Individual: x; Individual: y :: ys] =
  Sub2[p[atMost[y]];  p[atMost[x]];  p[y]] then AtMostStep[y; ys]
 end
 AtLeastStep[Individual: x; Individual: y :: empty] =
  Sub2[p[atLeast[x]]; p[x];          p[y]]
 end
 AtLeastStep[Individual: x; Individual: y :: ys] =
  Sub2[p[atLeast[x]]; p[atLeast[y]]; p[x]] then AtLeastStep[y;ys]
 end
in  GradedRels[S;T;p;Val;v0::v1::valS] 
and AtMostInitial[v0;v1] and AtMostStep[v1;valS]
and AtLeastStep[v0;v1::valS]
\end{lstlisting}
\begin{lstlisting}[morekeywords={let,in,given,EquivalentWith}]
ontology GradedRelsSub_Significance =
 GradedRels_Significance then
 GradedRelsSub[PhysicalObject; PhysicalObject; hasIngredient;
  Significance;[0Insignificant,1Subordinate,2Essential,3Dominant]]
\end{lstlisting}
%\end{multicols}    
    \caption{\ontology{GradedRelsSub} and instantiation of \ontology{GradedRelsSub\_Significance}.} 
    \label{fig:graded-rels-sub}
    \label{fig:graded-rels-insts}
\end{figure}

%\begin{figure}[t] 
%    \lstinputlisting[language=dolText,mathescape,morekeywords={Class, ObjectProperty,Domain,Range,SubPropertyOf,given,fit,let,in,Characteristics,
%    Reflexive,InverseOf,Transitive}]{Examples/GradedRelsLE_SignificanceExpanded.dol} 
%     \caption{Expansion of \ontology{GradedRelsLE\_Significance}.} 
%    \label{fig:graded-relsLE-expanded}
%\end{figure}

We can make use of the list constructor \texttt{::} to give 
different definitions for the same pattern according to 
the argument of the list parameter of that pattern.
This is a case distinction similar to pattern matching in
functional programming, that we call \emph{template matching} here
to avoid the overlap with ontology design patterns.
%At instantiation time, we go sequentially through 
In an instantiation, \Hets goes sequentially through the list of all definitions for a pattern
and checks whether the argument matches the parameter template.
%As soon as 
When a match is found, the body given in that definition is
used for instantiation.
If no match is found, the instantiation is incorrect.

%\begin{figure}[h!]
%% \begin{multicols}{2}
%%        \lstinputlisting[language=dolText,mathescape,morekeywords={Class, ObjectProperty,Domain,Range,SubPropertyOf,given,fit,let,in,Characteristics,
%%    Reflexive,InverseOf,Transitive}]{Examples/GradedRels_CrustStyle.dol}
%%    \columnbreak
%         \lstinputlisting[morekeywords={let,in,given,EquivalentWith}]{Examples/GradedRelsGE_Significance.dol}
% %   \end{multicols}
%    \caption{Instantiation of \ontology{GradedRelsGE}.} 
%    \label{fig:graded-rels-insts}
%\end{figure}

%\begin{figure}[h!]
%% \begin{multicols}{2}
%%    \columnbreak
%         \lstinputlisting[morekeywords={let,in,given,EquivalentWith}]{Examples/GradedRelsSub_Significance.dol}
% %   \end{multicols}
%    \caption{Instantiation of \ontology{GradedRelsSub}.} 
%\end{figure}

As an example, we provide a generic pattern for 
extending a sheaf of graded relation with subsumption relations,
%see Fig.~\ref{fig:graded-rels-le}. 
see Fig.~\ref{fig:graded-rels-sub}.
The idea is to introduce relations for expressing that a property
holds with at least or at most a grade, when the grades can be compared,
and to create a subsumption hierarchy between the relations 
\texttt{p\_G} and \texttt{p\_atLeast\_G}: if
a property \texttt{p} holds with a grade at least \texttt{G}, 
if it holds with grade \texttt{G} or it holds 
at least with a grade less than \texttt{G}. 
In this example, the recursion is shown both for a less-or-equal order 
(\texttt{atLeast}) and a greater-or-equal order (\texttt{atMost});
in the former, an initial step \texttt{AtMostInitial} is needed, 
while in the latter two cases for the recursion of \texttt{AtLeastStep} 
have to be distinguished to define a special final step for recursion termination.
When 
%\ontology{GradedRelsGE\_Significance}
\ontology{GradedRelsSub\_Significance} (Fig.~\ref{fig:graded-rels-insts})
has been expanded and  the names stratified, 
we obtain a %the following 
relation subsumption hierarchy between the graded relations
obtained %by instantiation of the pattern \texttt{GradedRels}
as follows (only the \texttt{atLeast} relations are shown):
%\ednote{Make sure that this stays on one page, once more text is available.}
%\ednote{withGrade\_Sig\_0Insignificant\_hasIngredient is
%  difficult to read. Better: has\_insignificant\_Ingredient.
%  Also: has\_at\_least\_insignificant\_Ingredient instead
%  of atLeastWith\_Sig\_0Insignificant\_hasIngredient.
%  MC: with the new convention for compound ids, we get much nicer names} is
\begin{verbatim}
 hasIngredient_atLeast_0Insignificant
  hasIngredient_0Insignificant
  hasIngredient_atLeast_1Subordinate
   hasIngredient_1Subordinate
   hasIngredient_atLeast_2Essential
    hasIngredient_2Essential
    hasIngredient_3Dominant
\end{verbatim}

% this is expansion of \ontology{GradedRelsLE\_Significance}
%\begin{verbatim}
%   hasIngredient
%    atMostWith_Sig_3Dominant-hasIngredient
%     with_Sig_3Dominant-hasIngredient
%     atMostWith_Sig_2Essential-hasIngredient
%      with_Sig_2Essential-hasIngredient
%      atMostWith_Sig_1Subordinate-hasIngredient
%       with_Sig_1Subordinate-hasIngredient
%       with_Sig_0Insignificant-hasIngredient
%\end{verbatim}

\section{Conclusions and Future Work}\label{sec:conclusions}

\cite{DBLP:conf/semweb/HitzlerGJKP17} introduces a list of desired capabilities
for a language for ODPs.
We list the desiderata, and how they are met by Generic \DOL{}, as follows:
\begin{enumerate}
\item \emph{Compatibility with the OWL standard and OWL supporting tools:} 

Ontologies generated with Generic \DOL{} are fully compatible with the OWL standard (after stratification). A Prot\'eg\'e plugin for Generic \DOL{} is planned.
\item \emph{Support for identification of ODPs as distinct from ontologies, and 
identification of relevant parts of ODPs:}

Patterns are generic (i.e. parameterized) ontologies; %, while ontologies are not parameterized. The 
the syntax %of generic ontologies 
allows a clear distinction between the parameters,
the imports and the body of a pattern.
\item \emph{Support for representing relevant relationships between patterns (refinement,
generalization etc.):}

%Generic \DOL{} supports generic views, which are 
The  \emph{views} in \DOL  can be used to define  
refinements between generic ontologies.
  %\ednote{This is not supported and this is also subject of future research, as also stated in \cite{DBLP:conf/semweb/HitzlerGJKP17}.} 
\item \emph{Support for identification of modules in ontologies generated using an ODP-based
approach:}

The module mechanisms in \DOL  allows the user to write
  ODPs in a way revealing their modular structure directly.
  %\ednote{This is very easy in the \DOL{} version of the ontology and in Hets, and not visible in the flat OWL version}.
\item \emph{Support for representing relationships between ontology modules and
the ODPs that have been used as templates for these modules:}

In Generic \DOL{}, this is the relation between a generic ontology and one of
  its instantiations. Hets displays this relation as a link in the
  development graph, which is a theory interpretation, semantically.
\item \emph{Extensibility of the language by means of community-provided patterns for
representing information about patterns and modules:}

Extensibility of the language is not supported (a higher-order extension of the language is under consideration). 
A structured repository and a  meta-ontology relating the GODPs in this repository are presently under development. 
\end{enumerate}

An important aspect is how to make the use of GODPs more intuitive for
ontology developers. A good GODP would have to provide
\begin{itemize}
\item a good choice of names for the pattern and for the parameters,
\item a documentation part informing the user  about the 
functionality of the pattern,
\item an instantiation example.
\end{itemize}
Ideally, working with GODPs will be done via a GUI that hides the body of the pattern from the
ontology developer (providing an appropriate documentation) and makes only those
parameters visible, which have to be instantiated.

\Hets{} support for the Generic \DOL{} language extensions introduced in this paper is currently in progress.

\bibliographystyle{unsrt}
\bibliography{paper}
\end{document}